\documentclass[sigconf]{acmart}

\AtBeginDocument{%
  \providecommand\BibTeX{{%
    \normalfont B\kern-0.5em{\scshape i\kern-0.25em b}\kern-0.8em\TeX}}}

\copyrightyear{2021}
\acmYear{2021}
\setcopyright{acmcopyright}\acmConference[BuildSys '21]{The 8th ACM International Conference on Systems for Energy-Efficient Buildings, Cities, and Transportation}{November 17--18, 2021}{Coimbra, Portugal}
\acmBooktitle{The 8th ACM International Conference on Systems for Energy-Efficient Buildings, Cities, and Transportation (BuildSys '21), November 17--18, 2021, Coimbra, Portugal}
\acmPrice{15.00}
\acmDOI{10.1145/3486611.3486645}
\acmISBN{978-1-4503-9114-6/21/11}

\usepackage{lineno}
\modulolinenumbers[5]
\usepackage{url}
\usepackage{amsmath}

\usepackage[ruled,vlined]{algorithm2e}

\usepackage{multirow}
\usepackage{comment}

\begin{document}
\title{Identifying the Relationship between Seasonal Variation in Residential Load and Socioeconomic Characteristics}


\author{Zhenyu Wang}
\affiliation{%
  \department{Faculty of Information Technology}
  \institution{Monash University}
  \city{Melbourne}
  \state{Victoria}
  \country{Australia}
  \postcode{VIC 3800}
}
\email{zwan149@student.monash.edu}

\author{Hao Wang}
\authornote{Corresponding author: \url{hao.wang2@monash.edu} (Hao Wang).}
\affiliation{%
  \department{Department of Data Science and AI}
  \department{and Monash Energy Institute}
  \institution{Monash University}
  \city{Melbourne}
  \state{Victoria}
  \country{Australia}
  \postcode{3800}
}
\email{hao.wang2@monash.edu}


\begin{abstract}

Smart meter data analysis can provide insights into residential electricity consumption behaviors. Seasonal variation in consumption is not well understood but yet important to utilities for energy pricing and services. 
This paper aims to develop a methodology to measure seasonal variations in load patterns and identify the relationship between seasonal variation and socioeconomic factors, as socioeconomic characteristics often have great explanatory power on electricity consumption behaviors. We first model the seasonal load patterns using a two-stage K-Medoids clustering and evaluate the relative entropy of the load pattern distributions between seasons. Then we develop decision tree classifiers for each season to analyze the importance of different socioeconomic characteristics factors. 
Taking real-world data as a case study, we find that income level is an essential factor influencing the pattern variation across all seasons. The number of children and the elderly is also a significant factor for certain seasonal changes.
\end{abstract}



\maketitle

\section{INTRODUCTION}
The installation of smart meters in residential premises has been growing explosively, and smart meter data recorded on an hourly basis or at minute intervals can improve the understanding of how electricity is consumed \cite{Albert2013}. Smart meter data analysis attracts electricity utilities and suppliers for developing better energy tariffs, programs, and services through load profiling \cite{Wei2021}. Establishing typical load patterns for consumers usually involves unsupervised learning methods like clustering \cite{Chicco2006}. Many works used clustering to find a single load pattern or load pattern distribution for consumers. However, the study in \cite{Azad2014} showed that seasonality is the key driver to more than one significant pattern of consumers. Several studies, such as \cite{Li2018,Guo2018}, also discovered seasonal variations in residential load, but the factors influencing such variations are less well studied. Understanding the variations in load patterns across seasons and key factors influencing seasonal variations is of great value to electricity utilities, retailers, and policymakers to design seasonal pricing schemes and evaluate the impact on different social groups. This paper aims to \textit{measure the seasonal variation in residential load patterns and identify the link between seasonal variation and socioeconomic factors.}

\subsection{Related Work and Motivation}
A review of related works shows the existence of seasonality in residential load patterns. In \cite{Guo2018}, load pattern shifts between seasons were found in households in China, and different households have different variations in their consumption behaviors. 
The study in \cite{Li2018} validated the importance of seasonality factor on both realistic and synthetic load profiles, where modeling seasonal variation can improve load prediction to some extent. The study in \cite{AsareBediako2014} suggested that the seasonal variation in load profiles should be handled in future smart grids management. However, existing studies lack insights into the driving or explanatory factors when analyzing seasonal variation in residential load. 
Therefore, this paper complements the existing studies in developing a methodology to identify the link between seasonal variation and socioeconomic factors.

Households characteristics have been used as features in analyzing load patterns, and some significant features are identified, though not for seasonal variations. For example, \cite{McLoughlin2012} used multiple linear regression models to evaluate the linkage between different socioeconomic factors and electricity consumption. The results indicated that characteristics like household age and social class have strong influences on the residential load. In \cite{Pombeiro2012}, a strong correlation has also been identified between socioeconomic factors and electricity consumption. The socioeconomic factors include consumers' income level, education level, and knowledge, determining consumers' attitudes on energy efficiency and thus changing their consumption behaviors. The study in \cite{Karatasou2019} found a positive association between socioeconomic status and residential load using a latent constructor approach. From these studies, we see a strong link between socioeconomic characteristics and consumers' habits and behaviors. This paper will show how socioeconomic factors are linked to and affect seasonal variation in load.

\subsection{Main Results and Contributions}
This paper develops methods to evaluate seasonal variation and study its link with socioeconomic factors. Specifically, we first develop a two-stage K-Medoids clustering method to model consumer load patterns based on our preliminary work \cite{Wang2021}.
Then, we quantify the seasonal variations by calculating the relative entropy of probability distributions of representative load patterns between adjacent seasons. Finally, by building classification models, we can establish the links between seasonal variations and socioeconomic factors and further analyze the importance of each factor. The key contributions of this paper are summarized as follows.

\begin{itemize}
  \item \textit{Extract Load Patterns}: We develop a two-stage K-Medoids method to cluster hourly load profiles on a daily basis, which can mitigate the impact of outliers and keep necessary load patterns for individual consumers.
  \item \textit{Quantify Seasonal Variations}: We use relative entropy, which measures the divergence between two probability distributions, as the metric to quantify the variation in load patterns between adjacent seasons.
  \item \textit{Link Socioeconomic Characteristics}: We build decision tree classifiers to estimate the variation of each seasonal change based on socioeconomic factors. We then analyze the effects according to the predictor importance.
  \item \textit{Generate insights using Real-world data}: We train our model using realistic data and observe that income level is the essential factor to determine whether consumers tend to vary their consumption behaviors when the season changes. The number of elderly residents and children are also important factors for some particular seasons.
\end{itemize}

\section{DATA DESCRIPTION}
This paper uses real-world data from the Pecan Street database \cite{pecan} to validate our methods. This database contains smart meter data of hourly load in kWh and socioeconomic information of households. After pre-processing the data, we obtain $417$ households as the consumers with the hourly consumption data over three years from $2015$ to $2017$ and corresponding socioeconomic factors. The socioeconomic factors include the number of residents in different age groups, annual income level, and education level.

\section{CONSUMER LOAD PATTERNS}
\begin{figure}[!t]
    \centering
    \includegraphics[width=0.9\linewidth]{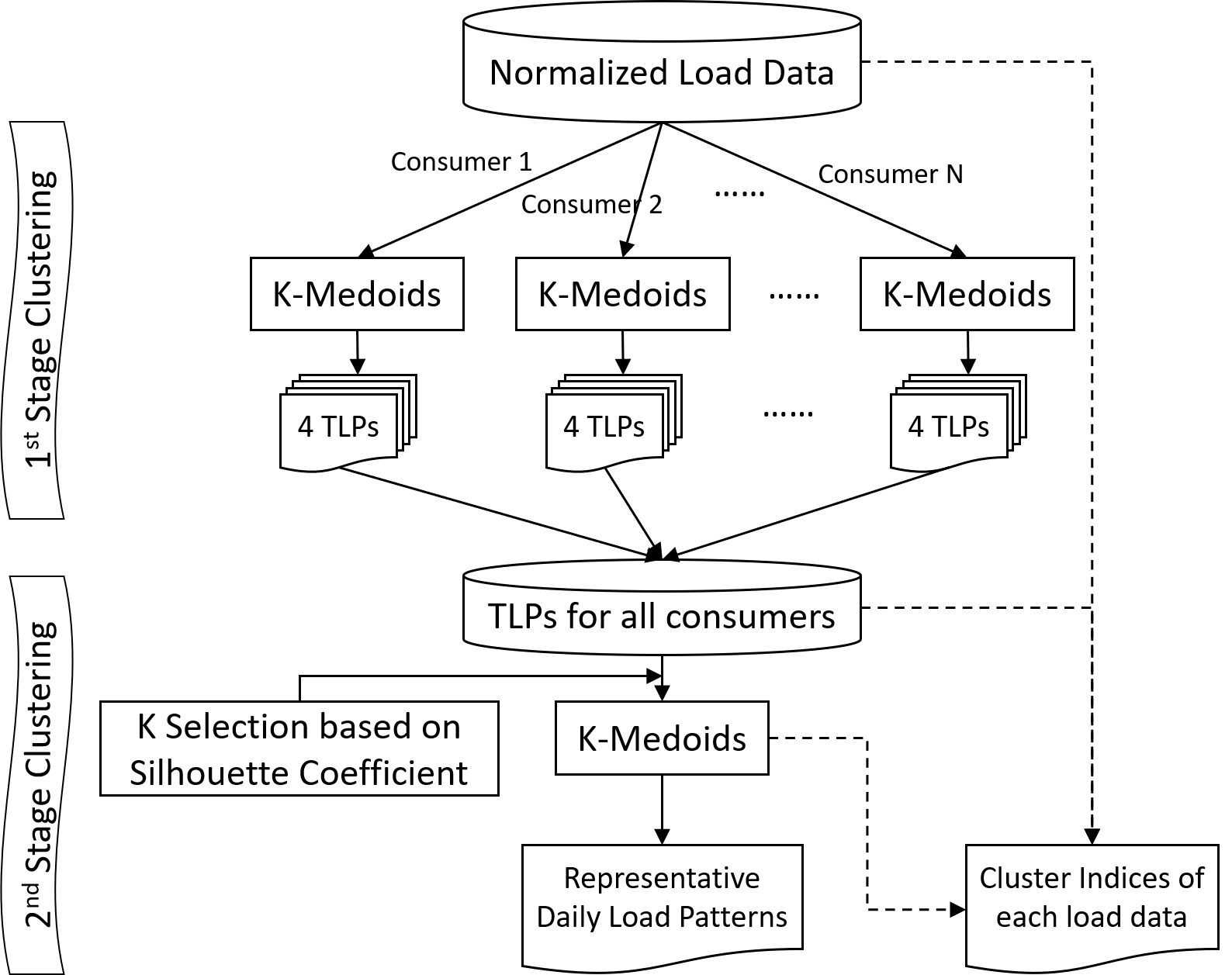}
    \vspace{-3.00mm} 
    \caption{Two-stage clustering framework.}
    \label{figure1}
    \vspace{-5.00mm}
\end{figure}

In this section, for the purpose of enhancing the interpretability of clusters and keeping sufficient patterns for each consumer at the same time, we develop a two-stage clustering method and extract a certain number of load profiles for every consumer before conducting the overall clustering for all consumers.
As shown in Figure \ref{figure1}, with the normalized residential load data, we adopt K-Medoids method \cite{Park2009} in the first-stage clustering to extract at most four typical load profiles (TLPs) for each consumer. In the second-stage clustering, we perform another K-Medoids clustering on the TLPs obtained from the first stage to produce representative daily load patterns among all consumers.

\subsection{Load Data Pre-processing}
The pre-processing of the residential load data includes removing outliers and normalization of load. The load data is denoted as $l_{d,t}^{n}$, where $n\in\mathcal{N}$ is the consumer index, $d$ is the date, and $t$ is the hour of a day. Since we are only interested in load variations instead of the magnitude of electricity consumption, we normalize the original load data $l_{d,t}^{n}$ to be $L_{d,t}^{n}$ in range of $[0,1]$ based on the minimum and maximum hourly load of consumer $n$ on day $d$.

\subsection{Two-stage Clustering}
To fit our model with real-world data that often contain outliers, we choose K-Medoids clustering \cite{Park2009} instead of the K-Means clustering and develop a two-stage clustering method. K-Medoids clustering uses the value of samples to be the centroids of clusters, which can help mitigate the impact of outliers and keep the original load shapes. In the first stage, we conduct K-Medoids clustering with $\hat{K}=4$ on each consumer to identify at most $4$ typical load patterns. We choose the number of clusters to be $4$, because we aim to evaluate pattern shifts over four seasons. Hence, we try to model each individual consumer's load patterns using at most $4$ patterns, i.e., one for each season. 
For the clustering outputs, we denote the set of four typical load profiles (TLPs) as $\hat{L}_n$ and the set of profile indices as $\hat{C}_n$ for each consumer $n$, shown as
\begin{equation}
  \hat{L}_n=\{\hat{L}_n^1,\hat{L}_n^2,\hat{L}_n^3,\hat{L}_n^4\}, 
  ~\hat{C}_n=\{\hat{C}_{d,t}^n, \forall d, \forall t\}, \label{clustering1}
\end{equation}
where $\hat{C}_{d,t}^n$ is the cluster index of which the load data instance $L_{d,t}^{n}$ is grouped.

After obtaining the four TLPs for each consumer, we apply another K-Medoids clustering to find representative daily load patterns among all consumers based on $\hat{L}=\{\hat{L}_n,n\in\mathcal{N}\}$ in the second stage. In other words, the second stage clustering is applied to the $4N$ TLPs obtained in the first stage. We denote the outputs from the second K-Medoids as
\begin{equation}
  \bar{L}=\{\bar{L}_1,\bar{L}_2,\ldots,\bar{L}_K\}, 
  ~ \bar{C}=\{\bar{C}_l, l\in \hat{L}\}, \label{clustering2}
\end{equation}
where $\bar{C}_l$ represents the cluster index of which the TLP $l$ is grouped.

Here, the number of clusters $K$ in the second K-Medoids will be selected based on the Silhouette coefficient \cite{Rousseeuw1987}, which measures the clustering performance. Specifically, it consists of two aspects: the cohesion factor $a_i$ as the mean distance between $i$ and other TLPs within the same cluster assigned for $i$ and the separation factor $b_i$ as the smallest average distance between $i$ and TLPs from the rest load clusters.

After the second K-Medoids clustering, we obtain the new cluster centroids and indices as stated in Eq. \eqref{clustering2}. Based on the first-stage clustering outputs in Eq. \eqref{clustering1}, we can update the cluster assignments for original normalized load data $L_{d,t}^{n}$ with final clusters in Eq. \eqref{clustering2}.

\section{SEASONAL VARIATION}
This section uses relative entropy to quantify and evaluate seasonal variations in residential load and then link seasonal variations to socioeconomic characteristics using classification models.

\subsection{Relative Entropy}
To further analyze the seasonal variation in load patterns, we need to compare the density of $K$ representative load patterns in each season. Relative Entropy (i.e., the Kullback-Leibler divergence \cite{Kullback1951}) can directly evaluate the divergence between two probability distributions, which makes it an appropriate measure in our analysis.

We define $D_{s,n}$ as the cluster index for a random day selected from season $s\in\left\{1,2,3,4\right\}$ for consumer $n$. The value of $s$ from $1$ to $4$ represents spring, summer, fall, and winter respectively. The sampling distribution of $D_{s,n}$ is calculated as:
\begin{equation}
  p(D_{s,n}=k)=I_{s,n}^k / I_{s,n},\label{cluster-distribution}
\end{equation}
where $I_{s,n}^k$ is the number of days that are classified into cluster $k$, and $I_{s,n}$ is the total number of days in season $s$ for consumer $n$.
Then, we calculate the relative entropy for the distributions of adjacent seasons $s$ and $s'$ as:
\begin{equation}
  RE(D_{s,n},D_{s',n})=\sum_{k=1}^{K}{p(D_{s',n}=k)\log_{K}{\frac{p(D_{s',n}=k)}{p(D_{s,n}=k)}}}, \label{re}
\end{equation}
where the log base is set to be the number of final clusters $K$ from the two-stage clustering.

In our case, higher $RE(D_{s,n},D_{s',n})$ indicates that the distribution of representative load patterns in season $s$ is less similar to that in season $s'$ for consumer $n$. This shows that consumer $n$ tends to vary the consumption behaviors when the season changes from $s$ to $s'$. Hence, the value of relative entropy is a measurement of the magnitude of variations in load patterns between different seasons.

\subsection{Classification Using Socioeconomic Factors}
Since socioeconomic characteristics could be the factors explaining consumers' adjustment of their consumption in different seasons, we attempt to use classification models to identify which factors are critical in deciding seasonal variations load patterns. We split the relative entropy into two classes based on its value comparing to a threshold, namely `variation' and `no variation'. We set the splitting threshold to be the relative entropy value for two distributions where one has $25\%$ difference from the other. 
The $25\%$ threshold is chosen based on the sample distribution, where two-thirds of samples are below this threshold. For predictors, we use consumer socioeconomic factors, including the number of residents in different age ranges, education level, and annual income level.

In our classification model, all the variables are often non-linear to the outcome, i.e., varying consumption habits or not. Therefore, we adopt decision tree classifiers to predict whether consumers have variations in load patterns between seasons using socioeconomic factors. We build separate classifiers for each season change, i.e., Spring-to-Summer, Summer-to-Fall, Fall-to-Winter, and Winter-to-Spring. We use the binary decision tree model and set the maximum number of splits as 20 to avoid overfitting. After training the models, we obtain the importance of each socioeconomic factor. If a factor has higher predictor importance, it indicates that the split on this factor will have a greater effect on the determination of the output. 

\section{Results and Discussion}

\subsection{Visualization of Seasonal Variation}
\begin{figure}[!b]
  \centering
  \vspace{-3.00mm} 
  \includegraphics[width=1\linewidth]{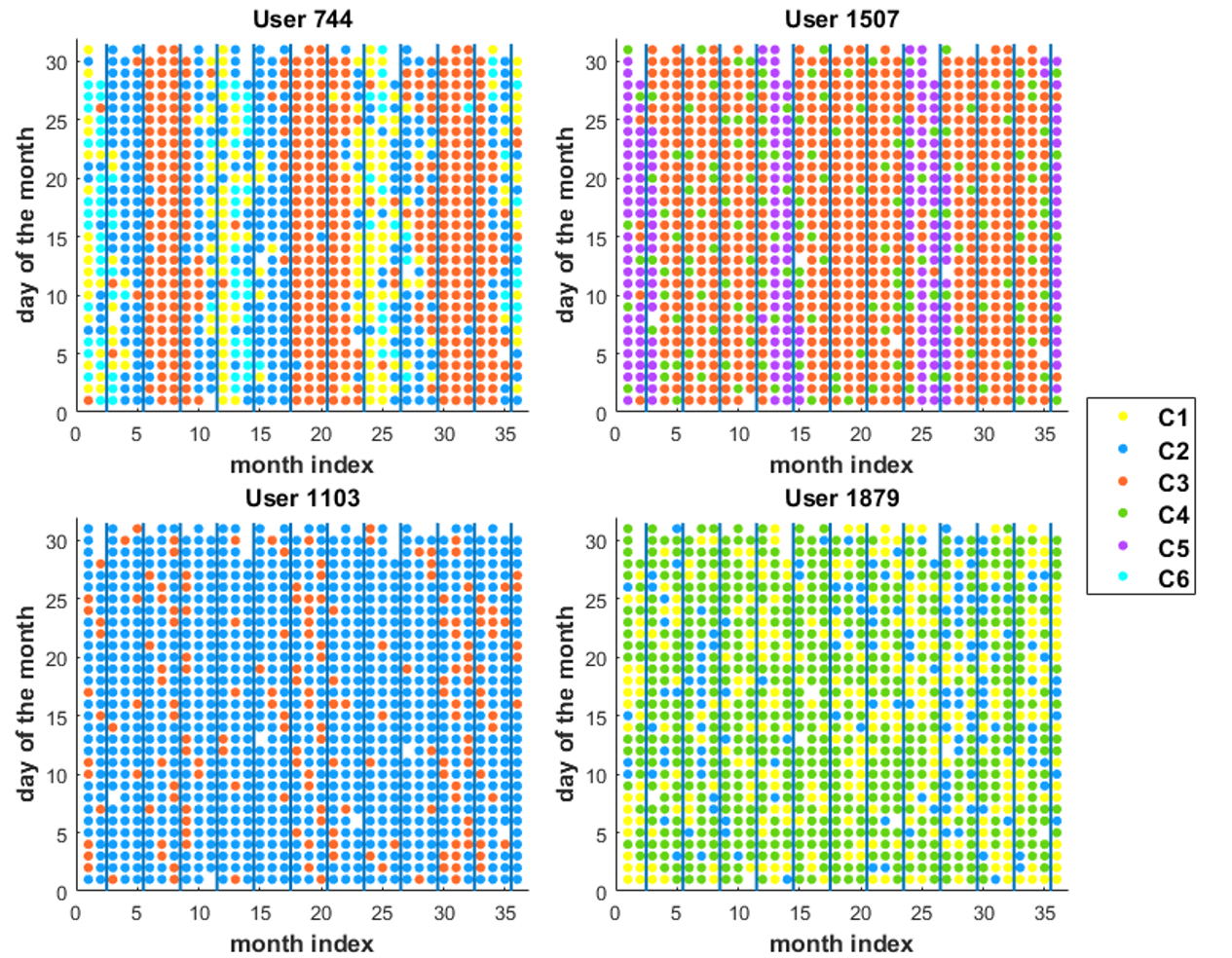}
  \vspace{-3.00mm} 
  \caption{Cluster Visualization for 4 selected consumers.}
  \label{figure4}
\end{figure}

After applying the two-stage K-Medoids clustering, we plot the distribution of representative load patterns throughout the dataset period of three years for four selected consumers as shown in Figure \ref{figure4}. Six representative clusters, i.e., C1-C6, are denoted with different colors. We can observe that `User 744' and `User 1507' have a clear seasonal variation in load patterns, comparing to `User 1103' and `User 1879'. Therefore, from these simple illustrations, we can find that environment parameters like weather and temperature change are not the only causes for seasonal variations in residential load patterns, and we refer to socioeconomic factors.

\begin{figure}[!t]
  \centering
  \includegraphics[width=0.9\linewidth]{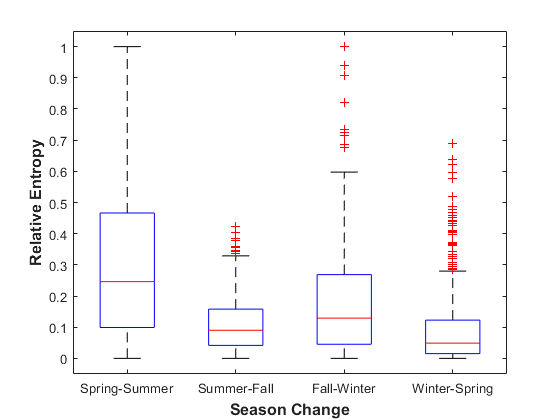}
  \vspace{-3.00mm}
  \caption{Box Plot of Relative Entropy for 4 Season Changes.}
  \label{figure5}
  \vspace{-3.00mm}
\end{figure}

After calculating the relative entropy to quantify the seasonal variation in load patterns, we obtain the box plot of the calculated relative entropy for four seasonal changes as shown in Figure \ref{figure5}. We can see that Spring-to-Summer has the largest median in relative entropy, which shows that residents are more likely to change their consumption habits when it turns hot in Summer. 

\begin{figure}[!b]
  \centering
  \vspace{-2.00mm} 
  \includegraphics[width=1\linewidth]{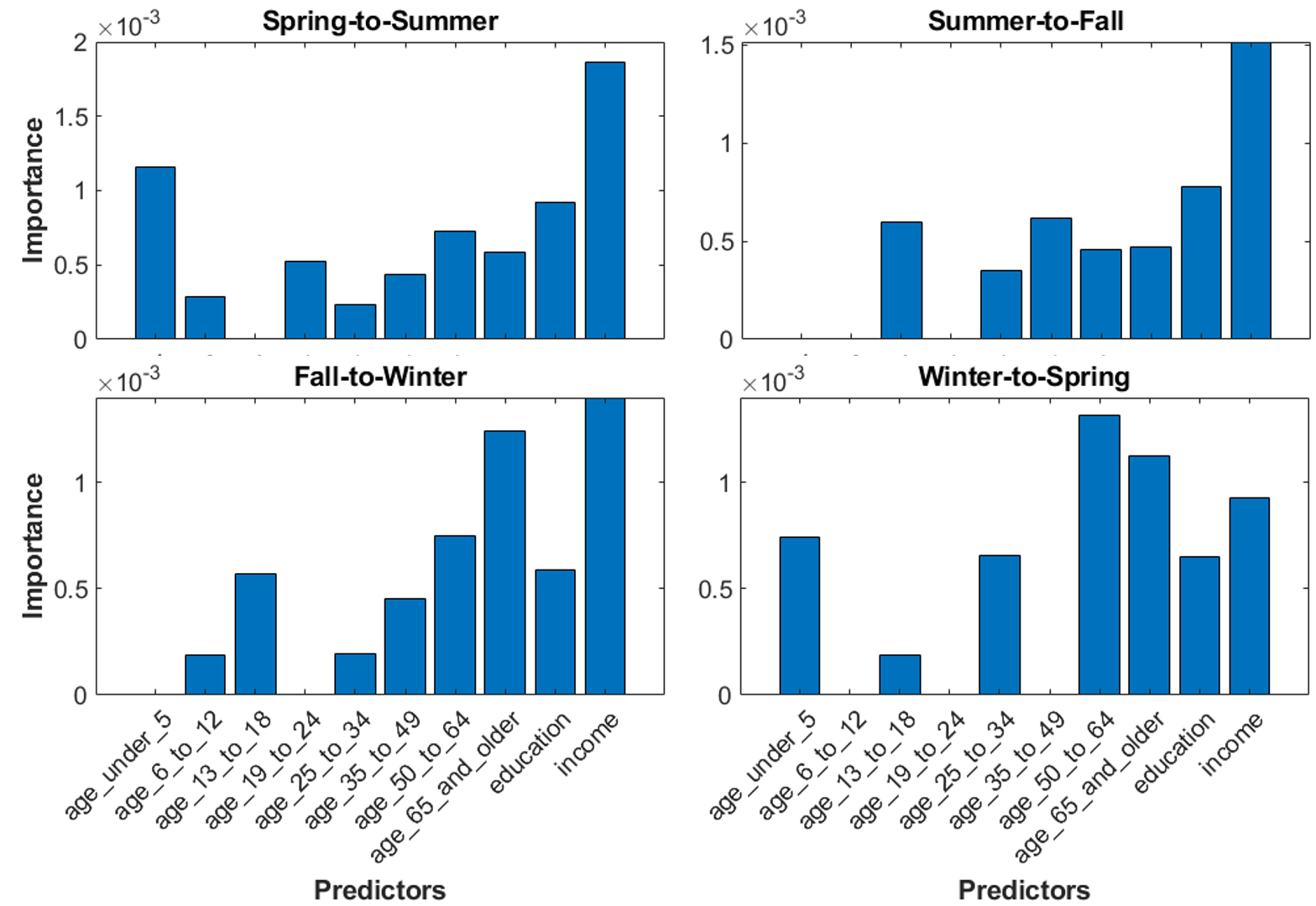}
  \vspace{-6.00mm} 
  \caption{Predictor Importance for 4 Classifiers.}
  \label{figure7}
\end{figure}

\subsection{Evaluation of Socioeconomic Factors}
We then train separate Decision Tree Classifiers for each change of seasons in a year using the available socioeconomic information with 5-fold cross-validation based on the relative entropy values. Figure \ref{figure7} plots the predictor importance in each classifier. We can observe that the income level is an important factor in all season changes when determining seasonal variation in load patterns, which is reasonable as families with higher incomes are more likely to have more diversity in electricity consumption. 
We also see that the number of children under the age of 5 has some importance in Spring-to-Summer and Winter-to-Spring models. This indicates that families with children tend to have different consumption habits in springtime. 
It is also noticeable that the number of elderly people over the age of 65 is another essential factor in Fall-to-Winter and Winter-to-Spring models. This shows that households with elderly people tend to have different consumption behaviors in wintertime.
From the results, the other predictors of different age ranges and education level do not have a distinct explanatory power in the determination of seasonal variation. This work can only investigate those showing higher importance in the designed model, and it will require further tests and control variable experiments to decouple different factors’ impact.

\section{Conclusion and Future Work}
This paper develops a method to analyze the relationship between socioeconomic characteristics and seasonal variations in residential load, which involves a two-stage K-Medoids clustering, relative entropy measurement, and Decision Tree classifiers. From a case study on real-world data, we find that income level is an essential factor to determine whether households tend to adjust their electricity consumption behaviors for different seasons. We also find that families with children and elderly people have variations in load patterns for particular season changes. Since we can only conjecture about the underlying mechanisms for these effects (e.g., higher temperature sensitivity among the elderly, seasonal allergies driving electricity use, etc.), this uncovered relationship merits further investigation. We acknowledge the limitations that this paper only provides some preparatory work, and the current results may be confined and short-sighted due to the specific model design.

Our future work aims to analyze different factors that may influence the seasonal variation in residential load with more datasets and evaluate the impact of seasonal electricity rates on different consumers.

\bibliographystyle{ACM-Reference-Format}
\bibliography{ref.bib}


\begin{thebibliography}{15}


\ifx \showCODEN    \undefined \def \showCODEN     #1{\unskip}     \fi
\ifx \showDOI      \undefined \def \showDOI       #1{#1}\fi
\ifx \showISBNx    \undefined \def \showISBNx     #1{\unskip}     \fi
\ifx \showISBNxiii \undefined \def \showISBNxiii  #1{\unskip}     \fi
\ifx \showISSN     \undefined \def \showISSN      #1{\unskip}     \fi
\ifx \showLCCN     \undefined \def \showLCCN      #1{\unskip}     \fi
\ifx \shownote     \undefined \def \shownote      #1{#1}          \fi
\ifx \showarticletitle \undefined \def \showarticletitle #1{#1}   \fi
\ifx \showURL      \undefined \def \showURL       {\relax}        \fi
\providecommand\bibfield[2]{#2}
\providecommand\bibinfo[2]{#2}
\providecommand\natexlab[1]{#1}
\providecommand\showeprint[2][]{arXiv:#2}

\bibitem[\protect\citeauthoryear{??}{pec}{2019}]%
        {pecan}
 \bibinfo{year}{accessed Oct. 1, 2019}\natexlab{}.
\newblock \bibinfo{booktitle}{\emph{Dataport}}.
\newblock Website, Pecan Street Inc.
\newblock
\newblock
\shownote{Available: https://www.pecanstreet.org/dataport/.}


\bibitem[\protect\citeauthoryear{Albert and Rajagopal}{Albert and
  Rajagopal}{2013}]%
        {Albert2013}
\bibfield{author}{\bibinfo{person}{Adrian Albert} {and} \bibinfo{person}{Ram
  Rajagopal}.} \bibinfo{year}{2013}\natexlab{}.
\newblock \showarticletitle{Smart Meter Driven Segmentation: What Your
  Consumption Says About You}.
\newblock \bibinfo{journal}{\emph{IEEE Transactions on Power Systems}}
  \bibinfo{volume}{28}, \bibinfo{number}{4} (\bibinfo{year}{2013}),
  \bibinfo{pages}{4019--4030}.
\newblock


\bibitem[\protect\citeauthoryear{Asare-Bediako, Kling, and
  Ribeiro}{Asare-Bediako et~al\mbox{.}}{2014}]%
        {AsareBediako2014}
\bibfield{author}{\bibinfo{person}{B. Asare-Bediako}, \bibinfo{person}{W.L.
  Kling}, {and} \bibinfo{person}{P.F. Ribeiro}.}
  \bibinfo{year}{2014}\natexlab{}.
\newblock \showarticletitle{Future residential load profiles: Scenario-based
  analysis of high penetration of heavy loads and distributed generation}.
\newblock \bibinfo{journal}{\emph{Energy and Buildings}}  \bibinfo{volume}{75}
  (\bibinfo{year}{2014}), \bibinfo{pages}{228--238}.
\newblock
\showISSN{0378-7788}


\bibitem[\protect\citeauthoryear{Azad, Ali, and Wolfs}{Azad
  et~al\mbox{.}}{2014}]%
        {Azad2014}
\bibfield{author}{\bibinfo{person}{Salahuddin~A Azad}, \bibinfo{person}{A~B
  M~Shawkat Ali}, {and} \bibinfo{person}{Peter Wolfs}.}
  \bibinfo{year}{2014}\natexlab{}.
\newblock \showarticletitle{Identification of typical load profiles using
  K-means clustering algorithm}. In \bibinfo{booktitle}{\emph{Asia-Pacific
  World Congress on Computer Science and Engineering}}.
  \bibinfo{publisher}{IEEE}, \bibinfo{address}{Nadi, Fiji},
  \bibinfo{pages}{1--6}.
\newblock


\bibitem[\protect\citeauthoryear{Chicco, Napoli, and Piglione}{Chicco
  et~al\mbox{.}}{2006}]%
        {Chicco2006}
\bibfield{author}{\bibinfo{person}{G. Chicco}, \bibinfo{person}{R. Napoli},
  {and} \bibinfo{person}{F. Piglione}.} \bibinfo{year}{2006}\natexlab{}.
\newblock \showarticletitle{Comparisons among clustering techniques for
  electricity customer classification}.
\newblock \bibinfo{journal}{\emph{IEEE Transactions on Power Systems}}
  \bibinfo{volume}{21}, \bibinfo{number}{2} (\bibinfo{year}{2006}),
  \bibinfo{pages}{933--940}.
\newblock


\bibitem[\protect\citeauthoryear{Guo, Zhou, Zhang, Yang, and Shao}{Guo
  et~al\mbox{.}}{2018}]%
        {Guo2018}
\bibfield{author}{\bibinfo{person}{Zhifeng Guo}, \bibinfo{person}{Kaile Zhou},
  \bibinfo{person}{Xiaoling Zhang}, \bibinfo{person}{Shanlin Yang}, {and}
  \bibinfo{person}{Zhen Shao}.} \bibinfo{year}{2018}\natexlab{}.
\newblock \showarticletitle{Data mining based framework for exploring household
  electricity consumption patterns: A case study in China context}.
\newblock \bibinfo{journal}{\emph{Journal of Cleaner Production}}
  \bibinfo{volume}{195} (\bibinfo{year}{2018}), \bibinfo{pages}{773--785}.
\newblock


\bibitem[\protect\citeauthoryear{Karatasou and Santamouris}{Karatasou and
  Santamouris}{2019}]%
        {Karatasou2019}
\bibfield{author}{\bibinfo{person}{S. Karatasou} {and} \bibinfo{person}{M.
  Santamouris}.} \bibinfo{year}{2019}\natexlab{}.
\newblock \showarticletitle{Socio-economic status and residential energy
  consumption: A latent variable approach}.
\newblock \bibinfo{journal}{\emph{Energy and Buildings}}  \bibinfo{volume}{198}
  (\bibinfo{year}{2019}), \bibinfo{pages}{100--105}.
\newblock
\showISSN{0378-7788}


\bibitem[\protect\citeauthoryear{Kullback and Leibler}{Kullback and
  Leibler}{1951}]%
        {Kullback1951}
\bibfield{author}{\bibinfo{person}{Solomon Kullback} {and}
  \bibinfo{person}{Richard~A. Leibler}.} \bibinfo{year}{1951}\natexlab{}.
\newblock \showarticletitle{On information and sufficiency}.
\newblock \bibinfo{journal}{\emph{The annals of mathematical statistics}}
  \bibinfo{volume}{22}, \bibinfo{number}{1} (\bibinfo{year}{1951}),
  \bibinfo{pages}{79--86}.
\newblock


\bibitem[\protect\citeauthoryear{Li, Allinson, and He}{Li
  et~al\mbox{.}}{2018}]%
        {Li2018}
\bibfield{author}{\bibinfo{person}{Matthew Li}, \bibinfo{person}{David
  Allinson}, {and} \bibinfo{person}{Miaomiao He}.}
  \bibinfo{year}{2018}\natexlab{}.
\newblock \showarticletitle{Seasonal variation in household electricity demand:
  A comparison of monitored and synthetic daily load profiles}.
\newblock \bibinfo{journal}{\emph{Energy and Buildings}}  \bibinfo{volume}{179}
  (\bibinfo{year}{2018}), \bibinfo{pages}{292--300}.
\newblock


\bibitem[\protect\citeauthoryear{McLoughlin, Duffy, and Conlon}{McLoughlin
  et~al\mbox{.}}{2012}]%
        {McLoughlin2012}
\bibfield{author}{\bibinfo{person}{Fintan McLoughlin}, \bibinfo{person}{Aidan
  Duffy}, {and} \bibinfo{person}{Michael Conlon}.}
  \bibinfo{year}{2012}\natexlab{}.
\newblock \showarticletitle{Characterising domestic electricity consumption
  patterns by dwelling and occupant socio-economic variables: An Irish case
  study}.
\newblock \bibinfo{journal}{\emph{Energy and Buildings}}  \bibinfo{volume}{48}
  (\bibinfo{year}{2012}), \bibinfo{pages}{240--248}.
\newblock
\showISSN{0378-7788}


\bibitem[\protect\citeauthoryear{Park and Jun}{Park and Jun}{2009}]%
        {Park2009}
\bibfield{author}{\bibinfo{person}{Hae-Sang Park} {and}
  \bibinfo{person}{Chi-Hyuck Jun}.} \bibinfo{year}{2009}\natexlab{}.
\newblock \showarticletitle{A simple and fast algorithm for K-medoids
  clustering}.
\newblock \bibinfo{journal}{\emph{Expert Systems with Applications}}
  \bibinfo{volume}{36}, \bibinfo{number}{2} (\bibinfo{year}{2009}),
  \bibinfo{pages}{3336--3341}.
\newblock


\bibitem[\protect\citeauthoryear{Pombeiro, André, and Carlos}{Pombeiro
  et~al\mbox{.}}{2012}]%
        {Pombeiro2012}
\bibfield{author}{\bibinfo{person}{Henrique Pombeiro}, \bibinfo{person}{Pina
  André}, {and} \bibinfo{person}{Silva Carlos}.}
  \bibinfo{year}{2012}\natexlab{}.
\newblock \showarticletitle{Analyzing residential electricity consumption
  patterns based on consumer’s segmentation}. In
  \bibinfo{booktitle}{\emph{Proceedings of the first international workshop on
  information technology for Energy Applications}}.
  \bibinfo{publisher}{CEUR-WS.org}, \bibinfo{address}{Lisbon, Portugal},
  \bibinfo{pages}{29--38}.
\newblock


\bibitem[\protect\citeauthoryear{Rousseeuw}{Rousseeuw}{1987}]%
        {Rousseeuw1987}
\bibfield{author}{\bibinfo{person}{Peter~J. Rousseeuw}.}
  \bibinfo{year}{1987}\natexlab{}.
\newblock \showarticletitle{Silhouettes: a graphical aid to the interpretation
  and validation of cluster analysis}.
\newblock \bibinfo{journal}{\emph{Journal of computational and applied
  mathematics}}  \bibinfo{volume}{20} (\bibinfo{year}{1987}),
  \bibinfo{pages}{53--65}.
\newblock


\bibitem[\protect\citeauthoryear{Wang and Wang}{Wang and Wang}{2021}]%
        {Wang2021}
\bibfield{author}{\bibinfo{person}{Zhenyu Wang} {and} \bibinfo{person}{Hao
  Wang}.} \bibinfo{year}{2021}\natexlab{}.
\newblock \showarticletitle{Analyzing Sea sonal Variation in Residential Load
  Patterns via Two-Stage Clustering and Relative Entropy: Poster}. In
  \bibinfo{booktitle}{\emph{The Twelfth ACM International Conference on Future
  Energy Systems (ACM e-Energy)}}. \bibinfo{publisher}{ACM},
  \bibinfo{address}{Torino, Italy}, \bibinfo{pages}{286–287}.
\newblock


\bibitem[\protect\citeauthoryear{Wei and Wang}{Wei and Wang}{2021}]%
        {Wei2021}
\bibfield{author}{\bibinfo{person}{Zhuo Wei} {and} \bibinfo{person}{Hao Wang}.}
  \bibinfo{year}{2021}\natexlab{}.
\newblock \showarticletitle{Characterizing Residential Load Patterns by
  Household Demographic and Socioeconomic Factors}. In
  \bibinfo{booktitle}{\emph{The Twelfth ACM International Conference on Future
  Energy Systems (ACM e-Energy)}}. \bibinfo{publisher}{ACM},
  \bibinfo{address}{Torino, Italy}, \bibinfo{pages}{244--248}.
\newblock


\end{thebibliography}

\end{document}